\documentclass{emulateapj}
\usepackage{apjfonts}

\slugcomment{Accepted for publication in ApJ (19 May)}

\shorttitle{Spitzer IRAC imaging of the 3C 273 Jet}

\shortauthors{Uchiyama et al.}

\begin{document}

\title{Shedding New Light on the 3C~273 Jet with 
the \emph{Spitzer Space Telescope}}

\author{Yasunobu~Uchiyama,\altaffilmark{1,6} C.~Megan~Urry,\altaffilmark{1} 
C.~C.~Cheung,\altaffilmark{2,3} Sebastian~Jester,\altaffilmark{4} 
Jeffrey~Van~Duyne,\altaffilmark{1} Paolo~Coppi,\altaffilmark{1} 
Rita~M.~Sambruna,\altaffilmark{5} Tadayuki~Takahashi,\altaffilmark{6,7}
Fabrizio~Tavecchio,\altaffilmark{8} and Laura~Maraschi\altaffilmark{8}  }

\altaffiltext{1}{Yale Center for Astronomy and Astrophysics, 
Yale University, 260 Whitney Avenue, New Haven, CT 06520;
uchiyama@astro.isas.jaxa.jp}
\altaffiltext{2}{Jansky Postdoctoral Fellow; National Radio Astronomy
Observatory}
\altaffiltext{3}{Kavli Institute for Particle Astrophysics and Cosmology, 
Stanford University, Stanford, CA 94305}
\altaffiltext{4}{School of Physics and Astronomy, Southampton University,
Southampton SO17 1BJ, United Kingdom}
\altaffiltext{5}{NASA/GSFC, Greenbelt, MD 20771}
\altaffiltext{6}{ISAS/JAXA, 3-1-1 Yoshinodai, Sagamihara, Kanagawa, 229-8510, Japan}
\altaffiltext{7}{Department of Physics, University of Tokyo, 7-3-1 Hongo, Bunkyoku, 
Tokyo 113-0033, Japan}
\altaffiltext{8}{Osservatorio Astronomico di Brera, 
via Brera 28,  20121 Milano, Italy}

\begin{abstract}
We have performed infrared imaging of the jet of the quasar 3C~273 
at wavelengths 3.6 and $5.8\ \mu\rm m$ 
with the Infrared Array Camera (IRAC) on the \emph{Spitzer Space Telescope}. 
When combined with the radio, optical and X-ray measurements, 
the IRAC photometry of the X-ray-bright jet knots clearly shows that 
the optical emission is dominated by the high-energy 
emission component of the jet, not by the radio synchrotron component, 
as had been assumed to date. 
The high-energy component, represented by a power-law 
from the optical through X-ray, may be due to a second synchrotron 
component or to inverse Compton scattering of ambient photons. 
In the former case,  we argue that  
  the acceleration of 
protons exceeding energies of $E_{\rm p} \sim 10^{16}\ \rm eV$ or 
possibly even to $E_{\rm p} \sim 10^{19}\ \rm eV$ 
would be taking place in the jet knots of 3C~273 assuming 
that the acceleration time is proportional to the particle gyroradius. 
In contrast, the inverse Compton model, into which 
highly relativistic Doppler beaming has to be incorporated, 
requires very low-energy electrons of 
$E_{\rm e} \sim 1\ \rm MeV$ in the jet knots. 
The present polarization data in the radio and optical would favor 
the former interpretation in the case of the 3C~273 jet. 
Sensitive and detailed measurements of optical polarization are important 
to establish the radiation mechanism responsible for the high-energy emission. 
The present study offers new clues as to the controversial origin of the X-ray emission seen in many quasar jets. 
\end{abstract}


\section{Introduction}
\label{sec:intro}

\begin{figure*}  
\epsscale{.80}
\plotone{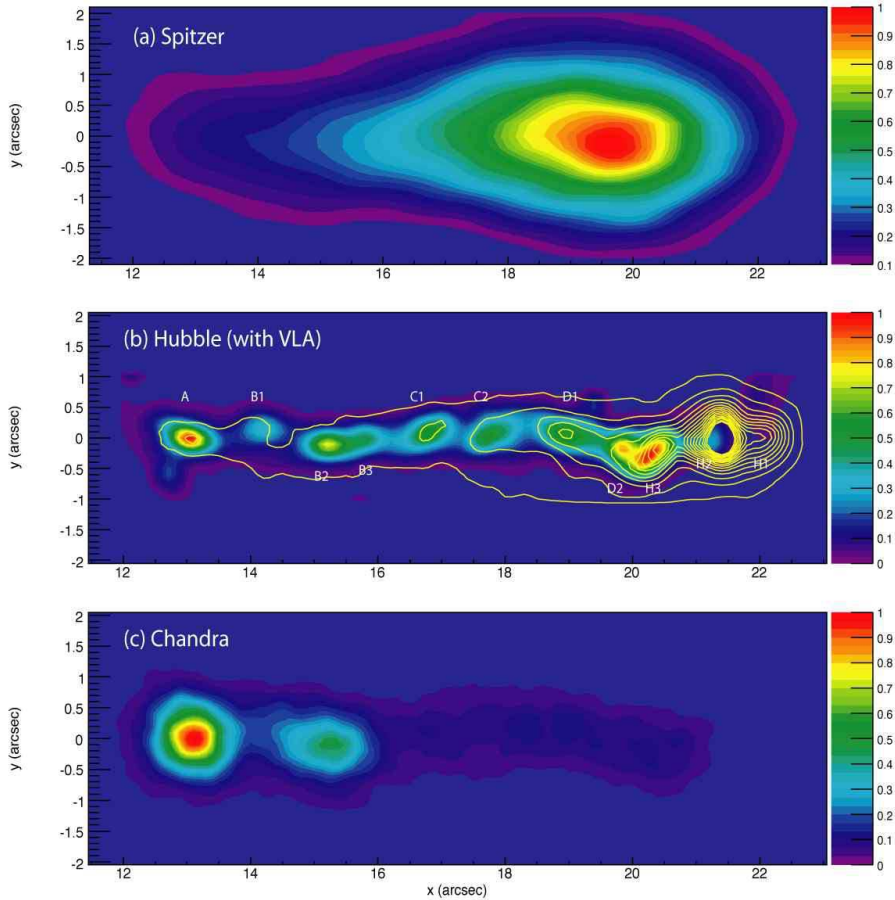}
\caption{Multifrequency images of the 3C~273 jet 
normalized to the the maximum pixel. 
({\it a}) \emph{Spitzer} IRAC (PSF-subtracted) image at $3.6\ \mu\rm m$;  
({\it b}) \emph{HST} image at 620 nm with VLA (at 2.0 cm) contours 
overlaid [adapted from \citet{Jes05}];  
({\it c}) \emph{Chandra} image in the 0.4--6 keV interval.
The origin of the coordinate system is at the 3C~273 core and the horizontal axis, 
parallel to the ridge line of the jet, is set at position angle $222\degr$
(the quasar is to the left at $x=0$). 
\label{fig:image} }
\end{figure*}

Large-scale jets extending to hundreds of kiloparsec distances from 
the quasar nucleus often radiate their power predominantly in X-rays. 
This has been revealed by surveys with the \emph{Chandra} X-ray Observatory 
 \citep{Sam04,Mar05} 
following its unexpected discovery of an X-ray jet in 
the quasar PKS 0637$-$752 
\citep[$z=0.65$:][]{Cha00,Sch00}. 
Despite  extensive observational and theoretical work, 
the dominant X-ray emission mechanism operating in quasar jets  remains 
unsettled \citep[e.g.,][]{AD04,Har06}.
Neither naive single-component synchrotron nor  synchrotron-self-Compton models
can fully  explain the spectral energy distributions (SEDs) at radio, 
optical and X-ray wavelengths. 
A widely discussed hypothesis for the strong X-ray emission is 
{\it relativistically enhanced} inverse Compton (IC)
scattering off the cosmic microwave background (CMB) photons \citep{Tav00,CGC01}. 
In this model, the bulk velocity of the jet is assumed to be highly relativistic 
all the way to nearly megaparsec distances, with a 
Doppler factor of $\delta \sim 10$.
Applying the beamed IC model to 
a large sample of \emph{Chandra}-detected quasar jets, 
 \citet{KS05} found  a range of 
 $4 \la \delta \la 16$ for $\sim 90\%$ of the sources. 

Alternative scenarios for producing X-ray emission at such large distances 
from the central engine include a variety 
of  ``non-conventional" synchrotron models. The X-ray emission 
could be explained by a 
spectral  hump  in a synchrotron spectrum as a result of  
reduced IC cooling in the Klein-Nishina regime provided that a Doppler factor 
of the jet is large \citep{DA02}.  Alternatively, 
a second synchrotron component responsible for the X-rays
could be formed by turbulent acceleration in the shear 
boundary layer \citep{SO02} or 
by hypothetical high-energy neutral beams (neutrons and gamma-rays) 
from the central engine \citep{AD01,Ner02}.
Finally, synchrotron radiation may be produced by very-high-energy protons, 
$E_{\rm p} \ga 10^{18}\ \rm eV$ \citep{Aha02}.
Discussion of these models is hampered mostly by the limited observational 
windows  available  so far. Apart from radio observations, only 
\emph{Hubble Space Telescope} (\emph{HST}) and \emph{Chandra} 
have the capability to resolve jets of powerful quasars.

To explore the mid-infrared properties of the extended jet 
emission from powerful quasars and thereby to shed new light on the riddle of the 
\emph{Chandra}-detected jets,  
we  performed \emph{Spitzer} IRAC imaging observations of four 
powerful quasars during the first observing cycle.  
The first result of this program, based on the observation of 
quasar PKS 0637$-$752 \citep{Uch05}, reported 
the first mid-infrared detection of  jet knots in a powerful quasar, and 
 demonstrated that  mid-infrared fluxes are of great 
importance in studying the SEDs, given the large gap between 
the radio and optical wavelengths.  
Following \citet{Geo05}, we also showed that 
in the context of the beamed IC model, observations with the \emph{Spitzer}  
IRAC can  constrain the matter content of jets \citep{Uch05}.

In this paper, we present \emph{Spitzer} IRAC 
results for the well-known jet in the nearby bright quasar 3C~273. 
We combine \emph{Spitzer} infrared measurements with available  data from the 
\emph{Very Large Array} (\emph{VLA}), \emph{HST}, and \emph{Chandra}, 
 forming 
an unprecedented data set to study the physics in extragalactic jets, 
 not  easily attainable for any of the other \emph{Chandra}-detected  jets found
in more distant quasars\footnote{A list of \emph{Chandra}-detected 
extragalactic jets is found at http://hea-www.harvard.edu/XJET/}.
With these data we show that the multiwavelength SED 
can be decomposed into two distinct  nonthermal  components.
Based on our finding, 
we consider plausible scenarios for the origin of the X-ray emission, and 
derive quantitative constraints on the physical state of the jet knots.

This paper is organized as follows. In \S\ref{sec:3c273} we summarize 
previous results on the 3C~273 jet. 
In \S\ref{sec:observation} 
we present analysis of the new \emph{Spitzer} IRAC  data  and describe 
the  radio, optical, and X-ray data for the same emission regions.  
The multiwavelength spectral energy distributions 
are presented and discussed in \S\ref{sec:results}.
Concluding remarks are given in \S\ref{sec:conclusions}.

The redshift of 3C~273 is $z=0.158$ \citep{Sch63}, so we 
adopt a luminosity distance of $D_L = 749\ \rm Mpc$ (at this distance $1\arcsec$ 
corresponds to a linear scale of  $2.7\ \rm kpc$) 
for a $\Lambda$CDM cosmology with
$\Omega_{\rm m}=0.27$, $\Omega_{\Lambda}=0.73$, and 
$H_0=71\ \rm km\ s^{-1}\ Mpc^{-1}$ \citep[e.g.,][]{Spe03}.
We define the spectral index, $\alpha$, in a conventional way according to the relation 
$f_{\nu} \propto \nu^{-\alpha}$, where $f_{\nu}$ is the flux density 
at frequency $\nu$.

\section{A Brief Summary of the 3C 273 Jet}
\label{sec:3c273}

The presence of the kiloparsec scale jet of 3C~273 was recognized already 
in the famous  quasar-discovery  paper of \citet{Sch63}.
The jet of 3C~273 has been extensively studied 
\citep[see][for reviews]{Cou98,Sta04}:  in the radio
\citep{DMC85,Con93}, in the near-infrared \citep{MR94,Neu97,Jes05}, 
in the optical \citep{RM91,Bah95,Jes01,Jes05}, 
and in the X-rays \citep{HS87,Roe00,Mar01,Sam01}.
Figure~\ref{fig:image} shows the jet at various wavelengths
(see \S\ref{sec:jetimages} for details).
Radio and optical emission from the large-scale jet are almost certainly 
synchrotron in nature 
given their measured polarizations and the coincidence in orientation. 
However, the emission mechanism giving rise to 
the X-rays has been controversial (see below).

The one-sided radio jet appears continuous all the way from the nucleus 
to the  ``head"  region 
increasing in brightness toward the head \citep{Con93}. 
The head region is located $\simeq 23\arcsec$  from the nucleus 
or a projected distance of 62 kpc.
The jet shows a sideways oscillation or  ``wiggle"  with 
respect to  the ridge line of the jet.
The nonthermal radio spectrum is characterized by spectral index 
$\alpha_{\rm radio} \simeq 0.7\mbox{--}0.8$; the spectral index shows little variation 
 along the jet. 
At the southeastern  side the jet is accompanied by an extended but narrow 
 ``lobe" that shows  a quite steep spectrum 
 of $\alpha_{\rm radio} \sim 1.5$ \citep{DMC85}. 
The faint lobe may be backflow from the hot spots  or 
may represent outer boundary layers. 

Optical imaging of the large-scale jet with \emph{HST} \citep{Bah95} 
revealed a structured morphology that is 
characterized by successive bright knots interspersed by  regions of weak emission.
The optical jet, extending from knot A to 
H2\footnote{In this paper, following \citet{Jes05}, 
 we adopt a nomenclature of knots as depicted in Fig.~\ref{fig:image}b. 
Note that there are many versions of the knot labeling 
in the literature.}, 
 is coincident with the radio features, but the brightness profile in the optical 
differs from that in the radio frequencies. 
The optical spectra are harder in regions close to the core \citep{RM91}, 
in such a way that 
spectral index declines from $\alpha_{\rm opt} \simeq 0.6$ at knot A 
to $\alpha_{\rm opt} \simeq 1.5$ at D2/H3 \citep{Jes01}.
It has been demonstrated that the flux at 300 nm (near ultraviolet) shows 
an excess over the extrapolation from lower frequencies \citep{Jes02}. 

Both the radio and optical jet emission are  linearly polarized. 
From $x=15\arcsec$ to $19\arcsec$ 
(where $x$ denotes the angular distance from the core)
the degree of radio polarization is 
$10\mbox{--}20\%$, and the average is consistent with the optical value 
\citep{RM91,RCM96}. 
This agreement argues 
 that both the optical and radio emission have a common (synchrotron) origin. 
The inferred magnetic field is longitudinal along the ridge line of the jet 
out to $x \simeq 21\arcsec$, beyond which 
the sudden transition of the magnetic field configuration from longitudinal to transverse 
is observed in the polarization data in both the radio \citep{Con93} and 
optical \citep{SR89,RM91}.
Based on the distinctive polarization pattern and the radiation spectrum, 
the head region
is widely considered to harbor  terminal  shocks and, as such, 
is  physically different from its upstream knots. 

Excess X-ray emission associated with the jet has been reported 
based on  observations made with  \emph{EINSTEIN} \citep{HS87}
and \emph{ROSAT} \citep{Roe00}.
\emph{Chandra} observations 
have confirmed the X-ray emission along the jet and allowed  detailed study 
of its spatial and spectral properties \citep{Sam01,Mar01}.
 X-ray flux is detected along the whole length of the optical jet\footnote{The presence of faint jet emission closer to the core ($x<10\arcsec$)  is known 
 in both the optical \citep{Mar03} and X-ray \citep{Mar01}.},  
from $x \simeq 12\arcsec$ (knot A) to $\simeq 21\arcsec$. Regions closer to the 
core, specifically knots A and B2, are particularly bright in the X-ray. 
The X-ray spectrum shows a power-law continuum, 
with  spectral index $\alpha_{\rm x} \simeq 0.6\mbox{--}0.9$, 
indicative of its nonthermal nature \citep{Mar01}.
For knot A, a synchrotron-self-Compton (SSC) model (with an equipartition 
magnetic field and without Doppler beaming) predicts 
the SSC X-ray flux lower by three orders of magnitude than the observed flux 
\citep[see][]{Roe00,Sam01,Mar01}.

It has been claimed that 
the innermost knot A has a straight power law continuum extending 
from the radio to X-ray, while the subsequent knots are  
described by a power law with a cut-off around $10^{14}$ Hz plus 
a separate X-ray component \citep{Roe00,Mar01}.
If so, the radio through X-ray emission in knot A may be ascribed to 
 synchrotron emission from the same electrons as the radio emission; 
the X-rays in the other knots are then produced in some other way. 
On the other hand,  \citet{Sam01} argued 
 that knot A as well as subsequent knots has a spectral break at optical 
wavelengths and suggested the SED is comprised of two components, 
 a synchrotron part (radio to optical) and a beamed IC part (X-ray and beyond). 
Finally, based on deep \emph{VLA} and \emph{HST} observations,  
\citet{Jes02} found a significant flattening in the near-infrared to ultraviolet 
spectra and argued that the radio-optical emission cannot be ascribed to 
single-population synchrotron radiation requiring an additional high-energy
component \citep[see also][]{Jes05}.
Below we construct the SEDs of the jet knots based on the extensive 
\emph{VLA}, \emph{Spitzer}, \emph{HST}, and \emph{Chandra} data,
which allow us to elucidate the controversy in interpreting the shapes of 
the broadband SEDs. 
We then discuss possible origins for the high-energy emission, X-rays in particular.

\section{Observations and Analysis}
\label{sec:observation}

\subsection{Spitzer IRAC Observation}

\begin{figure}  
\epsscale{1.0}
\plotone{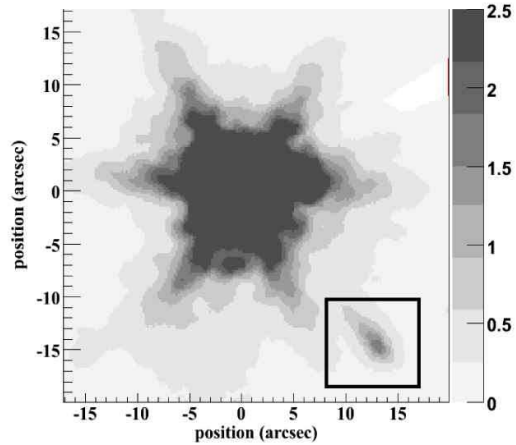}
\caption{\emph{Spitzer} IRAC image of 3C~273 and its jet 
at $3.6\ \mu\rm m$  in units of 
$\rm MJy\ sr^{-1}$. North is up and east is to the left. The origin of 
the coordinates is at the quasar core. 
The jet emission is clearly visible in the box drawn in the lower right quadrant.  
\label{fig:rawimage} }
\end{figure}

The infrared observations of  3C~273 with \emph{Spitzer} IRAC were carried out 
on 2005 June 10 as part of our GO-1 program (\emph{Spitzer} Program ID 3586).
IRAC is equipped with a four-channel camera,  
InSb arrays at wavelengths 3.6 and $4.5\ \mu\rm m$ and 
Si:As arrays at  5.8 and $8.0\ \mu\rm m$, 
each of which has a field of view of $5\farcm2\times 5\farcm2$ \citep{Faz04}.
The InSb arrays operate at $\sim 15\ \rm K$ while 
the Si:As arrays operate at $\sim 6\ \rm K$.
Only the pair of 3.6 and $5.8\ \mu\rm m$ arrays,  observing 
the same sky simultaneously,  was chosen 
for our observation, to obtain longer exposures in one pair 
of bandpasses as opposed to two pairs with half the exposure time.
The 3.6/5.8 pair was chosen because of its better spatial resolution and sensitivity.  
The pixel size in both arrays is $\simeq 1\farcs22$. The point-spread 
functions (PSFs) are characterized by  widths of 
$1\farcs66$ and $1\farcs88$ (FWHM) for the 
3.6 and $5.8\ \mu\rm m$ bands, respectively \citep{Faz04}. 
The flux measurement of point sources with IRAC is calibrated to an 
accuracy of 2\% \citep{Rea05}.

For each IRAC band, we obtained 26 well-dithered  frames,  each with a 30 s 
integration time.
The pipeline process (version S12.0.2) at the \emph{Spitzer} Science Center (SSC) 
 yielded 26 calibrated  images.
These calibrated frames  were combined into a mosaic image with 
a pixel size of $0\farcs2$ 
using the SSC package MOPEX \citep{MM05}, which removes 
 spurious sources such as cosmic rays  and moving objects 
 based on inter-frame comparisons. 
 The dithered frames were combined using the ``drizzling" technique \citep{FH02}.
Finally, we made use of the astrometric positions of three field optical sources 
that have infrared counterparts 
to accurately ($\la 0\farcs2$) locate the position of 3C~273 and its jet in the 
 IRAC image.
The $3.6\ \mu\rm m$  mosaic image is shown 
in Fig.~\ref{fig:rawimage}. Note that although the quasar core is extremely bright, 
the jet emission clearly stands out to the southwest. 

As is evident from Fig.~\ref{fig:rawimage}, the infrared emission of the jet 
is contaminated by the PSF wings of the bright quasar core. 
Therefore,  measuring the infrared fluxes of the jet components 
requires careful removal of the wings of the core underneath the jet emission. 
Unlike the case of PKS 0637$-$752 \citep{Uch05}, 
the point-spread function provided 
by the SSC is not applicable for the  much longer jet of 3C~273. 
Instead we made use of deep IRAC images of a very bright star in the 
Extended Chandra Deep Field South (E-CDFS) to generate a PSF template 
that is bright enough out to a distance of $\sim 20\arcsec$ away from the core. 
In fact, the template star is brighter than 3C~273 itself 
by a factor of $\simeq 13$ in the $3.6\ \mu\rm m$ band and by a factor of 
$\simeq 4$ in the $5.8\ \mu\rm m$ band, respectively. 
Due to the saturation in both the 3C~273 and E-CDFS images, and possible 
contributions from the host galaxy of 3C~273, 
the PSF subtraction can only be carried out reliably for $x>10\arcsec$, 
but fortunately it covers the entire length of the main body 
of the optical and X-ray jet.
The PSF subtraction procedure introduces an additional systematic error
in the infrared flux determination. 
Based on the small level of intensity due to a residual in PSF subtraction
in the vicinity of the jet, we have estimated the systematic errors to be 
$0.05\ \rm MJy\ sr^{-1}$ for the $3.6\ \mu\rm m$ channel and 
$0.3\ \rm MJy\ sr^{-1}$ for the $5.8\ \mu\rm m$ channel, which are 
added in quadrature to the noise of each pixel when photometry is performed. 
The jet is so bright that 
 photometry errors due to noise become unimportant with respect to these 
systematic errors.

The core emission of 3C~273 itself is  
saturated in both the $3.6\ \mu\rm m$ and $5.8\ \mu\rm m$ science frames. 
An additional frame was taken in a 
high-dynamic-range mode with a short integration time (1 s), such that 
3C~273 itself is unsaturated; 
we use this frame to estimate the mid-infrared flux of the quasar core. 
We obtain the flux density of  the 3C~273 core to be 
$\simeq 170$ mJy at $3.6\ \mu\rm m$ 
and $\simeq 230$ mJy at $5.8\ \mu\rm m$, respectively. 
At the relevant wavelengths, the spectrum of 3C~273 is known to 
have a steady small bump at $\sim3.5\ \mu\rm m$ 
superposed on a variable power-law continuum \citep{Rob86}.
At the L band ($3.45\ \mu\rm m$),  
\citet{Tur99} list the mean flux density of $\bar{f_{\nu}} = 160.5\ \rm mJy$ 
with a dispersion (due to time variability) of $\sigma_{\nu} = 33.5\ \rm mJy$, 
with which the IRAC flux is consistent. 
The spectroscopic observation of 3C~273 on 2004 January 6 made with 
the Infrared Spectrograph (IRS) on \emph{Spitzer} gives the flux density 
of $f_{\nu} \simeq 190\ \rm mJy$ at $5.8\ \mu\rm m$ 
\citep{Hao05}.  
The power-law continuum in June 2005 (our observation) 
seems slightly brighter than January 2004. 
It should be noted, however, that the flux estimate from a 
 single short frame suffers from larger systematic errors 
as compared to the science frames.

\subsection{Multiwavelength Jet Images}
\label{sec:jetimages}

Figure \ref{fig:image}a shows the IRAC image at $3.6\ \mu\rm m$  
of the 3C~273 jet after subtraction of the PSF wings of the quasar core, where 
a line in position angle $222\degr$ from the core is adopted 
as a reference horizontal axis.  The $5.8\ \mu\rm m$ image is essentially similar 
to the $3.6\ \mu\rm m$ image but with slightly worse resolution. 
In Fig.~\ref{fig:image}, we also show high-resolution 
radio (at a wavelength of 2.0 cm obtained with \emph{VLA}),  
optical (at 620 nm with \emph{HST}), 
and X-ray (with \emph{Chandra}) images.
The \emph{VLA} and \emph{HST} images are same as 
those presented in \citet{Jes05}, both of which have an effective 
resolution (FWHM) of $0\farcs3$. (The resolution in the \emph{HST} image 
is artificially degraded from its native resolution to facilitate a meaningful 
comparison with other bands. The strongest radio spot H2 is truncated 
in the \emph{VLA} contour map.)
The \emph{Chandra} image is based on new and deep 
observations with the ACIS-S detectors (PI: Jester). We describe the data 
analysis of the \emph{Chandra} observations later in \S\ref{sec:chandra}
[see also Jester et al.\ (2006, submitted) for complementary analyses].
Though the X-ray image presented here is essentially 
identical to Fig.~1 of \citet{Mar01}, 
which was derived from the early calibration observations with the grating detectors, 
the spectral parameters are updated taking into account calibration improvement.

The IRAC infrared emission traces almost perfectly the main body of the optical jet;
the onset of the infrared jet coincides with the innermost optical 
knot A centered at $x \simeq 13\arcsec$, and 
the infrared jet starts fading around $x \simeq 21\arcsec$ 
in the head region. 
A wiggling shape with a transverse amplitude of $\Delta y\sim 0\farcs2$ 
that is evident in the optical can barely be seen in the $3.6\ \mu\rm m$ image.
The infrared jet brightness  shares an overall spatial pattern with the radio 
map in a sense that the brightness of the jet  increases  toward 
the outer part almost monotonically. 
However, the radio intensity peaks at H2 where the infrared jet has already 
started getting fainter. 
The overall jet in the IRAC bands looks more similar to that at 
the near-infrared wavelength \citep[see][for a $K^\prime$ band image]{Neu97}. 
In contrast, in the optical at 620 nm each knot is more or less of similar brightness, 
suggesting a spectral change from longer wavelengths. 

To further investigate the jet emission quantitatively, particularly 
in terms of multiwavelength analysis, 
we derive the flux densities of \emph{individual} knot components, 
at  radio, infrared, optical, and X-ray wavelengths. 
The results of the photometry are summarized in Table \ref{tbl-1}, and presented 
in a conventional $\nu f_\nu$ form in Figs.~\ref{fig:SED} and \ref{fig:SED2}.
In the subsequent sections, we describe the methods of our measurements 
in each band before proceeding with multiwavelength analysis in 
\S\ref{sec:results}.

\subsection{Infrared Photometry}
\label{sec:IRphoto}

\begin{figure}  
\epsscale{1.0}
\plotone{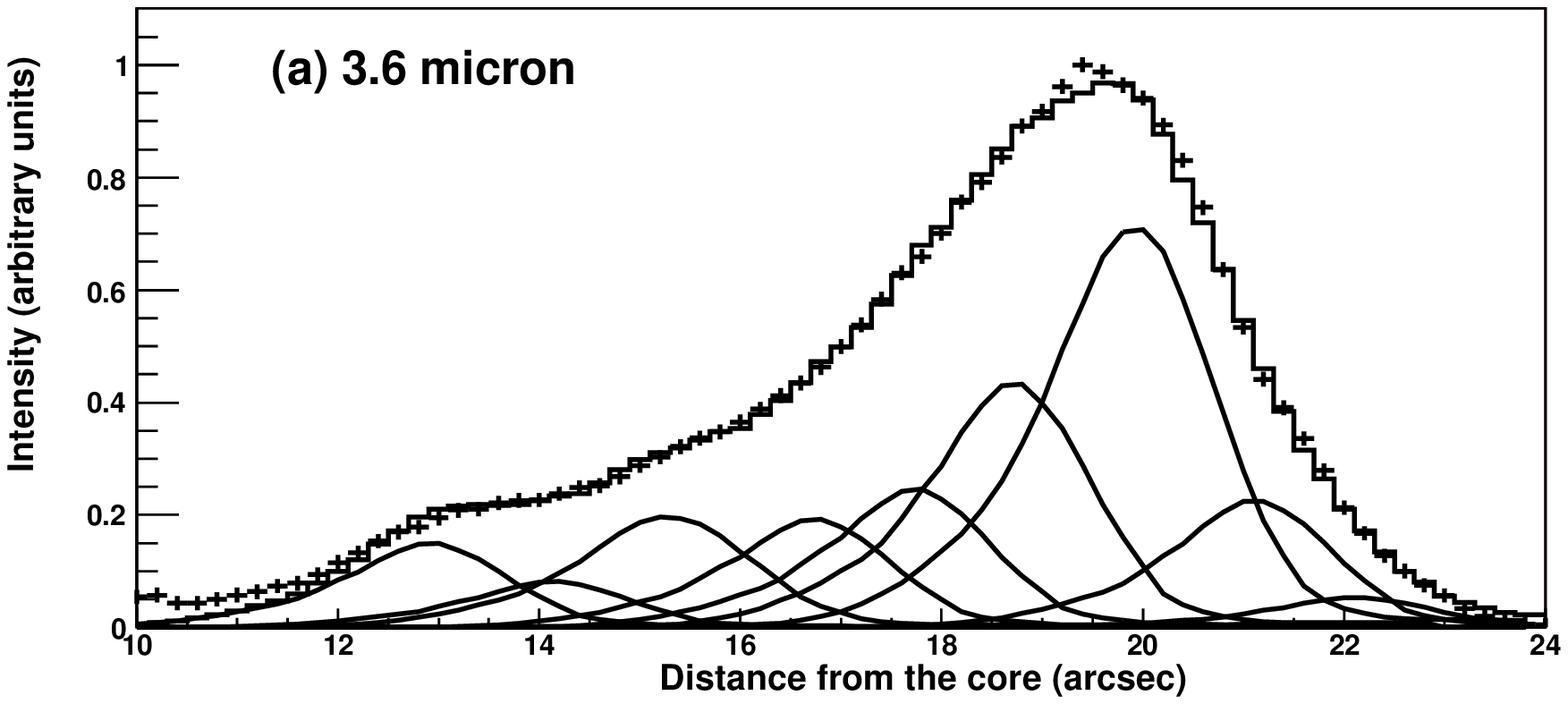}
\plotone{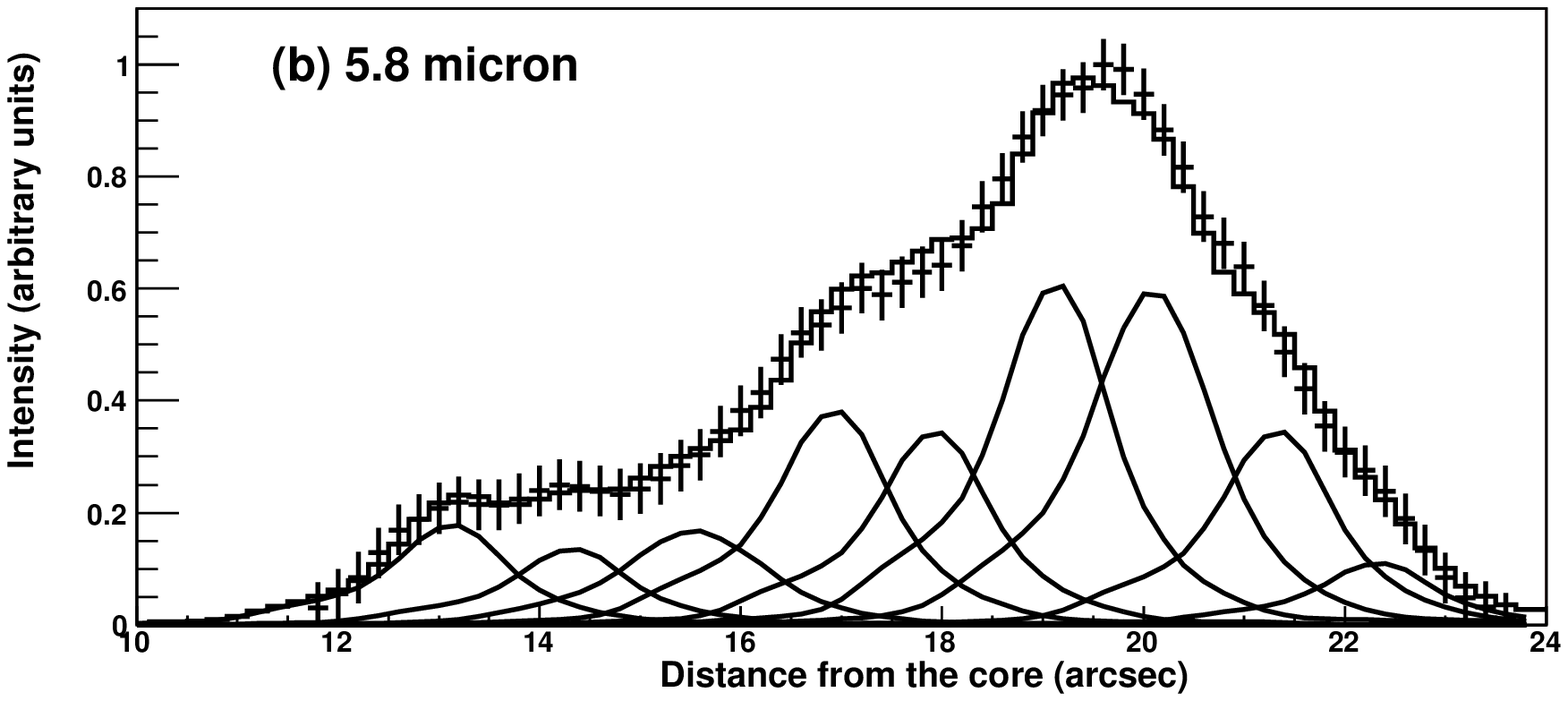}
\caption{Intensity profiles of the 3C~273 jet along the ridge line 
integrated over the vertical dimension from $-1\arcsec$ to $1\arcsec$
({\it a}) at $3.6\ \mu\rm m$  and  ({\it b}) at $5.8\ \mu\rm m$ 
obtained with \emph{Spitzer} IRAC, 
 together with the best-fit models comprised of 9 knot features. 
\label{fig:profile} }
\end{figure}

\begin{deluxetable*}{crrrrrrrrr}[ht]
\tablecolumns{10}
\tabletypesize{\footnotesize}
\tablecaption{Flux Densities of 3C~273 Jet Knots\label{tbl-1}}
\tablewidth{0pt}
\tablehead{
\multicolumn{1}{c}{} &
\multicolumn{9}{c}{Flux Density, $f_{\nu}$ }\\
\cline{2-10}\\ 
\colhead{Frequency} 
& \colhead{A} & \colhead{B1} & \colhead{B2/B3} 
&  \colhead{C1} & \colhead{C2} & \colhead{D1} 
& \colhead{D2/H3} & \colhead{H2}& \colhead{H1} \\
\colhead{$\nu$ (Hz)} & \multicolumn{9}{c}{}
}
\startdata
\sidehead{{\it VLA}\tablenotemark{a} (mJy):} 
$8.33\times10^{9}$ 
& 52  & 42 & 77 & 70 & 93 & 250 & 500 & 750 & 330 \\
$1.50\times10^{10}$ 
& 33 & 27 & 54 & 48 & 63 & 160 & 310 & 440 & 200 \\
$2.31\times10^{10}$ 
& 24 & 20 & 38 & 36 & 45 & 110 & 220 & 280 & 130 \\
\sidehead{{\it Spitzer}\tablenotemark{b} ($\mu$Jy):} 
$5.23\times10^{13}$ 
& 45$\pm$10 & 35$\pm$12 & 49$\pm$11 & 98$\pm$11
& 89$\pm$12 & 154 & 161 & 87$\pm$12 & 28$\pm$11 \\
$8.45\times10^{13}$ 
& 27 & 15$\pm$2.6 & 39 & 36
& 46 & 80 & 140 & 41 & 10$\pm$2.6 \\
\sidehead{{\it HST}\tablenotemark{c} ($\mu$Jy):} 
$1.87\times10^{14}$ 
& 7.6 & 4.4 & 11 & 9.0 & 11 & 21 & 30 & 5.4 & 1.5 \\
$4.85\times10^{14}$ 
& 3.6 & 1.7 & 4.0 & 2.6 & 2.5 & 3.6 & 6.2 & 1.1 & \nodata \\
$1.00\times10^{15}$ 
& 2.4 & 0.93 & 2.0 & 1.1 & 0.90 & 1.2 & 2.1 & 0.40 & \nodata \\
\sidehead{{\it Chandra}\tablenotemark{d} (nJy):} 
$2.14\times10^{17}$ 
& 46 & 11 & 26 & 5.8 & 5.1 & 6.1 & 7.6 & \nodata & \nodata \\
$3.77\times10^{17}$ 
& 28 &  6.2 & 14 & 2.7 & 2.6 & 3.7 & 3.9 & \nodata & \nodata \\
$1.05\times10^{18}$ 
& 12 & 2.7$\pm$0.3 & 5.1 & 0.9$\pm$0.2 
& 1.1$\pm$0.2 & 0.9$\pm$0.2 & 1.4$\pm$0.3 & \nodata & \nodata \\
\enddata
\tablecomments{The 90\% errors are shown only when exceeding $10\%$.
Otherwise a formal error of 10\% is assigned to conservatively account for 
the systematic errors (see the text). The optical fluxes are corrected for 
Galactic extinction using $A_V = 0.07$.
The X-ray fluxes are corrected for Galactic absorption using 
$N_{\rm H} = 1.7\times 10^{20}\ \rm cm^{-2}$.
}
\tablenotetext{a}{Nominal wavelengths are $\lambda =$ 3.6, 2.0 and 1.3 cm.}
\tablenotetext{b}{Nominal wavelengths are $\lambda = $ 5.73 and 3.55 $\mu$m.}
\tablenotetext{c}{Nominal wavelengths are $\lambda =$ 1.6, 0.62, 
and 0.3 $\mu$m. 
Knot  H1 is detected in the 1.6 $\mu$m band, but not at shorter 
wavelengths \citep{Jes05}.}
\tablenotetext{d}{Nominal energies are $\varepsilon =$ 0.89, 1.56, and 4.35 keV.}
\end{deluxetable*}

The angular separation of adjacent knots (typically $\simeq 1\arcsec$) is 
comparable to the width of the IRAC PSF, which allows us to derive knot fluxes 
nearly individually even though the infrared images do not show 
discernible knots.
We have modeled the two-dimensional brightness distribution of the whole 
length of the infrared jet with a series of PSFs of variable intensity centered on 
the positions of the known optical knots.
We constructed the PSF based on nearby template stars. To account for 
possible slight variations in the widths of the PSFs by  $\sim 10\%$, we adjusted 
each PSF width so that the PSF shape fits well the jet in the $y$-direction. 
As shown in Fig.~\ref{fig:image}b,  
11 knots are predefined from the optical image\footnote{
We use a term ``knot"  for labeling but faint inter-knot emission as well as 
the so-called cocoon are included in our photometry.}.
 Among them,  B2 and B3 were combined 
into a single component since the two knots are too close 
(separated only by $0\farcs5$) to be fit individually. 
Similarly, D2 and H3 were united. 
Elongated PSFs were applied to the combined knots so as to account for the 
multiplicity. 
Whereas the optical spectra of D2 and H3 are similar to one another, 
the optical slopes of B2 and B3 are slightly  different \citep{Jes05}; 
one must  therefore be  careful when interpreting the summed spectra
of B2/B3. 
Also, we note that an infrared-bright star,  St~2 in \citet{Jes05}, would 
contribute about 5\% of the flux density of knot D1. 
To summarize,  we fit the jet images 
at 3.6 and $5.8\ \mu\rm m$ with a series of 9 point-like components 
at fixed positions,  after convolving with the IRAC PSF. 

The PSF fit gives the flux density for each of the 9 components (Table~\ref{tbl-1}).
 Note that in determining the photometric errors  we have included 
 the uncertainties imposed by background 
(the wings of the core and the host galaxy) subtraction. 
Also,  photometry is subject to systematic errors due to  different 
spectral shapes,  as the standard IRAC calibration is strictly valid for the specific case 
of $f_{\nu} \propto \nu^{-1}$. 
Though we do not correct for this effect, the flux errors in this regard are 
less than 1\% for the spectral slopes of  interest. 

Figure \ref{fig:profile}  displays the one-dimensional profile of the infrared jet 
emission along the ridge line of the jet. 
The  measured brightness profile and the best-fit 9-component model 
are shown,  
integrated over the vertical dimension from $-1\arcsec$ to $1\arcsec$, 
together with its decomposition into individual PSFs. 

\subsection{Radio and Optical Photometry}

\begin{figure}  
\epsscale{1.0}
\plotone{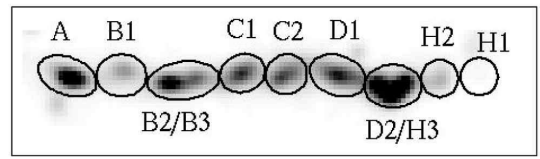}
\caption{Apertures used for radio and optical photometry, superposed on 
the \emph{HST} jet image (same as Fig.~\ref{fig:image}b).
\label{fig:regions} }
\end{figure}

We also performed photometry at three radio wavelengths 
(\emph{VLA}: 3.6, 2.0, and 1.3 cm) and three optical wavelengths 
(\emph{HST}: $1.6\ \mu\rm m$, 620 nm, and 300 nm).
The data sets used for the photometry are identical to those in \citet{Jes05}.
We re-analyzed the photometric images because \citet{Jes05} performed 
photometry with a $0\farcs3$-aperture  while we need 
knot-by-knot photometry to be directly compared with the infrared analysis. 
Unlike the infrared photometry described above,  a PSF-fitting procedure 
is not mandatory since the knot separation of $\sim 1\arcsec$ is large 
compared to the spatial resolution.
Therefore,  defining regions as depicted in Fig.~\ref{fig:regions}, 
we extracted the corresponding radio and optical 
fluxes  inside each aperture region. 
The optical fluxes are corrected for Galactic extinction ($A_V=0.07$); 
correction factors used here are 1.01 for  $1.6\ \mu\rm m$, 
 1.05 for 620 nm, and 1.12 for 300 nm, respectively. 
Photometric errors are small ($< 5\%$) in all cases for 
both the radio and optical measurements. 
Possible systematic errors 
such as those associated with deconvolution in the  interferometric radio data
or flux contamination from adjacent knots, though difficult to quantify, 
would be several percents. 

The morphologies of the knots in the jet are similar at all wavelengths; 
only the relative intensities of the knots change with frequency. 
The only exception to this  is knot B1, in which the radio emission does not have 
a corresponding feature to the optical knot. This issue should be kept in mind 
when interpreting the spectrum of knot B1.

\subsection{X-ray Photometry} 
\label{sec:chandra} 

Data used for the X-ray  photometry  presented here are based on 
four \emph{Chandra} ACIS-S observations performed 
between November 2003 and July 2004 with 
Observation IDs (Obs IDs) from 4876 to 4879.
The aimpoint was placed at the S3 CCD. 
The basic data reduction was done by using the CIAO software version 3.2.2 
and the CALDB 3.1.0 with the standard screening criteria applied. 
Periods of background flaring often caused by the increased flux of 
charged particles in orbit  have been excluded from each data set 
(e.g.,  a $\simeq 6\ \rm ks$ flare was found 
in the data of Obs ID 4879). 
After the screening processes 
a total exposure of 136 ks remains. 
We define three energy bands as 0.6--1 keV (soft), 1--2 keV (medium), 
and 3--6 keV (hard). 
The nominal energy $\varepsilon_{\rm norm}$ for each band is then 
the mean energy weighted by the effective area:  $\varepsilon_{\rm norm}=$ 
 0.89 keV, 1.56 keV, and 4.35 keV, for the soft, 
medium, and hard band, respectively. 

We have constructed  X-ray flux images in the three bands 
with a pixel size of $0\farcs2$ by dividing a counts map by an exposure map.
 These flux images, which are actually the weighted mean of four observation sets, 
  have  units of photons $\rm cm^{-2}\ s^{-1}\ keV^{-1}$ per pixel 
and allow us to perform X-ray  photometry.
Calculations of the exposure maps were performed 
with the CIAO tool {\it mkexpmap} assuming an incident spectrum of 
a power law with spectral index $\alpha = 1$, correcting for 
 the contamination buildup in the ACIS detectors. 
The results are insensitive to $\alpha$ since the corrections for 
different indices are minimal at the effective-area-weighted mean energy. 
Then we have measured  the flux densities at  $\varepsilon_{\rm norm}$ 
by integrating over 
relevant apertures and multiplying each image by $\varepsilon_{\rm norm}$. 
A similar but different procedure was adopted by \citet{Har04} 
to obtain the flux densities of the jet in 3C~120. 
Compared to the radio and optical cases (Fig.~\ref{fig:regions}), 
we elongated the object apertures in the transverse $y$-direction 
($|y| \leq  1\arcsec$) to accumulate photons such that 
the portion of photons lying outside the aperture should be $< 10\%$. 
We estimated the background using regions outside the jet in the vicinity of the 
extraction apertures. The background amounts only to $\la 2\%$ of the source flux 
for knot A even in the hard band where it has the largest contribution. 
Finally, to correct for the interstellar absorption of 
$N_{\rm H} = 1.7\times 10^{20}\ \rm cm^{-2}$ \citep{Alb93}, 
we  multiplied the measured fluxes by a factor of 1.08 and 1.02
in the soft and medium energy bands, respectively. 
The flux densities of the knot regions determined in this way are given 
in Table~\ref{tbl-1}. Most of the measurements have small statistical uncertainties 
less than 10\%. 
The jet emission downstream of knot B2 does not show well discernible peaks 
but emission features associated with knots C1, D1, and D2/H3 can be 
 recognized in the X-ray image, which justifies to some extent 
our aperture photometry for the downstream knots.

In addition to the photometry-based spectra, 
we have performed the more usual spectral analysis as a  consistency check. 
The X-ray spectra in the 0.5--6 keV range of the four observations 
were simultaneously fitted by a power law 
with a fixed absorption column $N_{\rm H} = 1.7\times 10^{20}\ \rm cm^{-2}$.
The resultant spectral index for knot A is found to be $\alpha = 0.84\pm0.04$.
Also, the index derived for a combined downstream region from knot C1 to 
D2/H3 is $\alpha = 1.06\pm0.07$. 
The photometry-based indices, on the other hand, are 
$\alpha \simeq 0.8$ and $\alpha \simeq 1.0$ for 
knot A and the combined downstream, respectively. 
Both methods are thus consistent with one another. 
It should be noted that we obtained somewhat steeper spectra, 
 by $\Delta\alpha \sim 0.2$, compared to the results presented 
by \citet{Mar01}. The difference arises most likely as a result of  absorption by 
the molecular contaminants in the ACIS detectors. The presence of such 
contaminants was not known at the time of  the \citet{Mar01} analysis 
 and consequently their  spectral analyses inevitably 
led to a  systematically  lower $\alpha$. 
Similarly, \citet{PW05} reanalyzed early \emph{Chandra} data on the  M~87 jet 
and found that contamination of the ACIS detector can account for 
the flatter X-ray spectra  than expected based on the optical slope as reported 
by \citet[][see their erratum]{WY02}.

\begin{figure*}[ht]
\epsscale{0.6}
\plotone{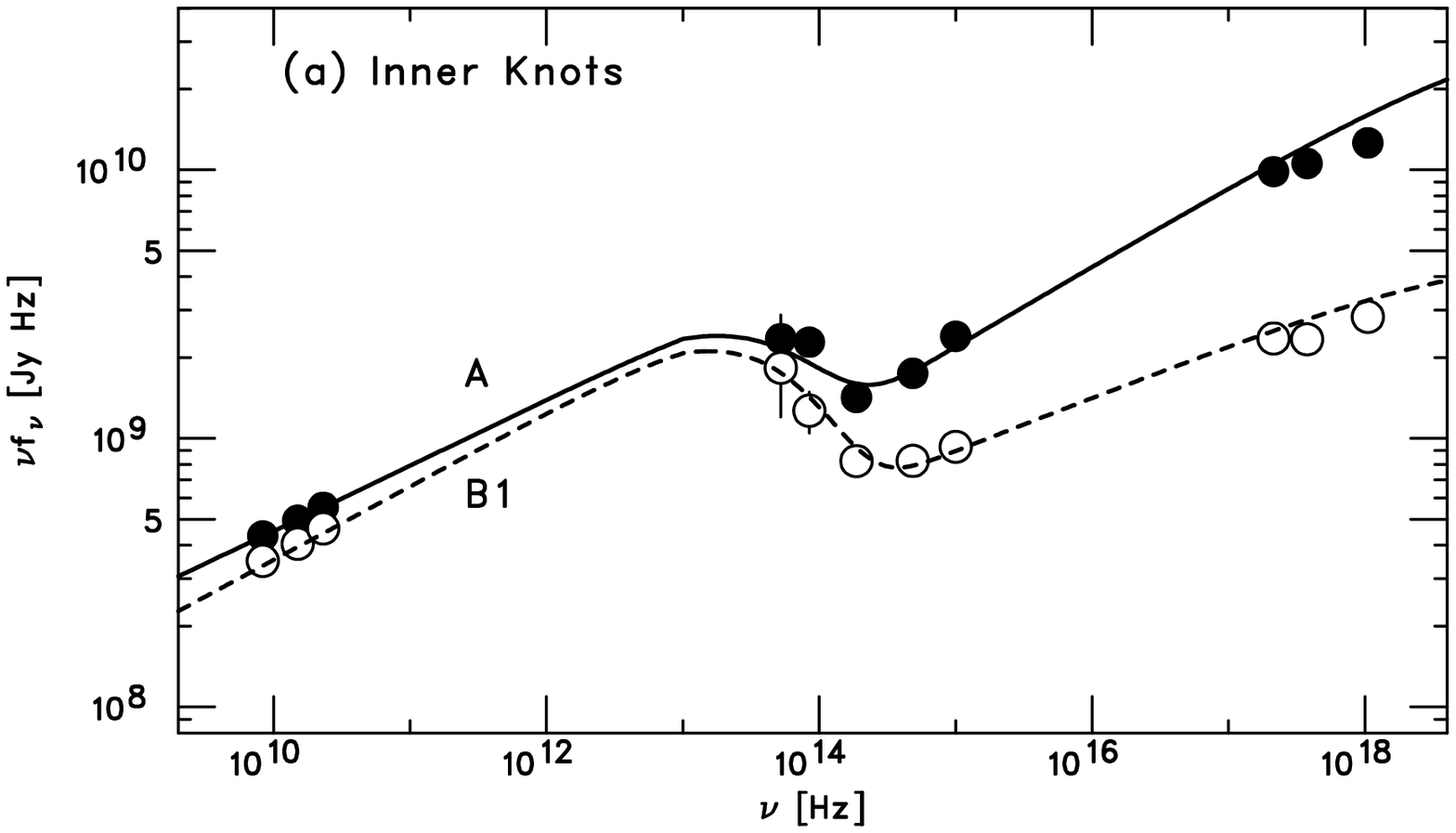} 
\plotone{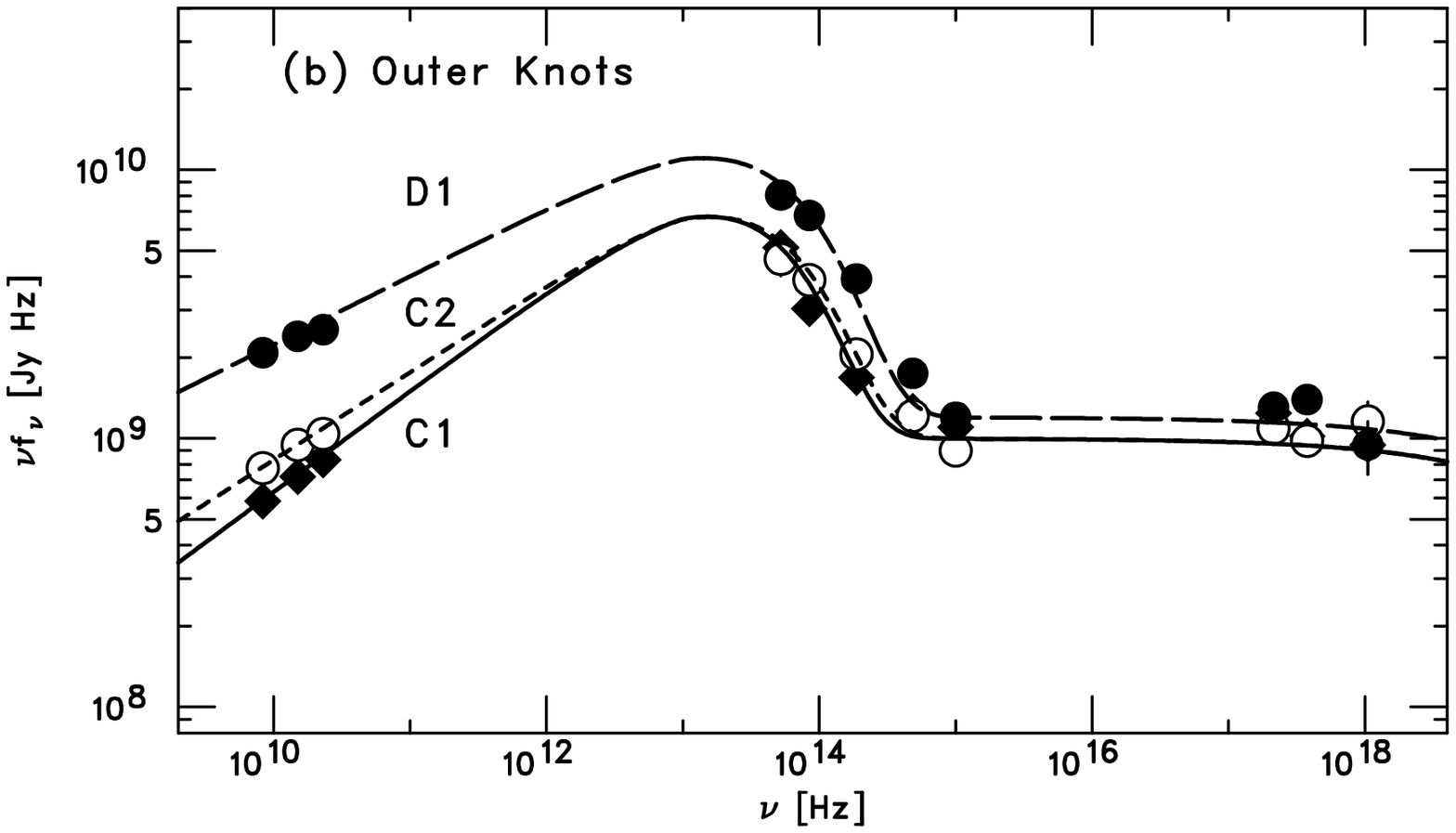}
\caption{Broadband SEDs of the 3C~273 large-cale jet. 
({\it a}) Inner knots A (filled circles) and B1 (open circles); ({\it b}) 
outer knots C1 (diamonds), C2 (open circles), and D1 (filled circles). 
The photometric data are given in Table \ref{tbl-1}. 
The curves represent the two-component phenomenological  model,  
namely equation (\ref{eq:fnu}):
$f_{\nu} = 
\kappa_1 \nu^{-\alpha_1} \exp\left[ -\left( \nu/\nu_{c} \right)^{a}\right] + 
\kappa_2 \nu^{-\alpha_2}$ with parameters in Table~\ref{tbl-2}.
Note that $1\ \rm Jy\, Hz = 10^{-23}\ erg\ cm^{-2}\ s^{-1}$. 
Figure~\ref{fig:IR-opt} zooms in the mid-infrared to ultraviolet part, 
where the photometric points are relatively cramped. 
\label{fig:SED}}
\end{figure*}

\section{Results and Discussion} 
\label{sec:results}

\subsection{SED of Jet Knots: Two-Component Nature} 
\label{sec:SED}

\begin{figure*}
\epsscale{0.9}
\plottwo{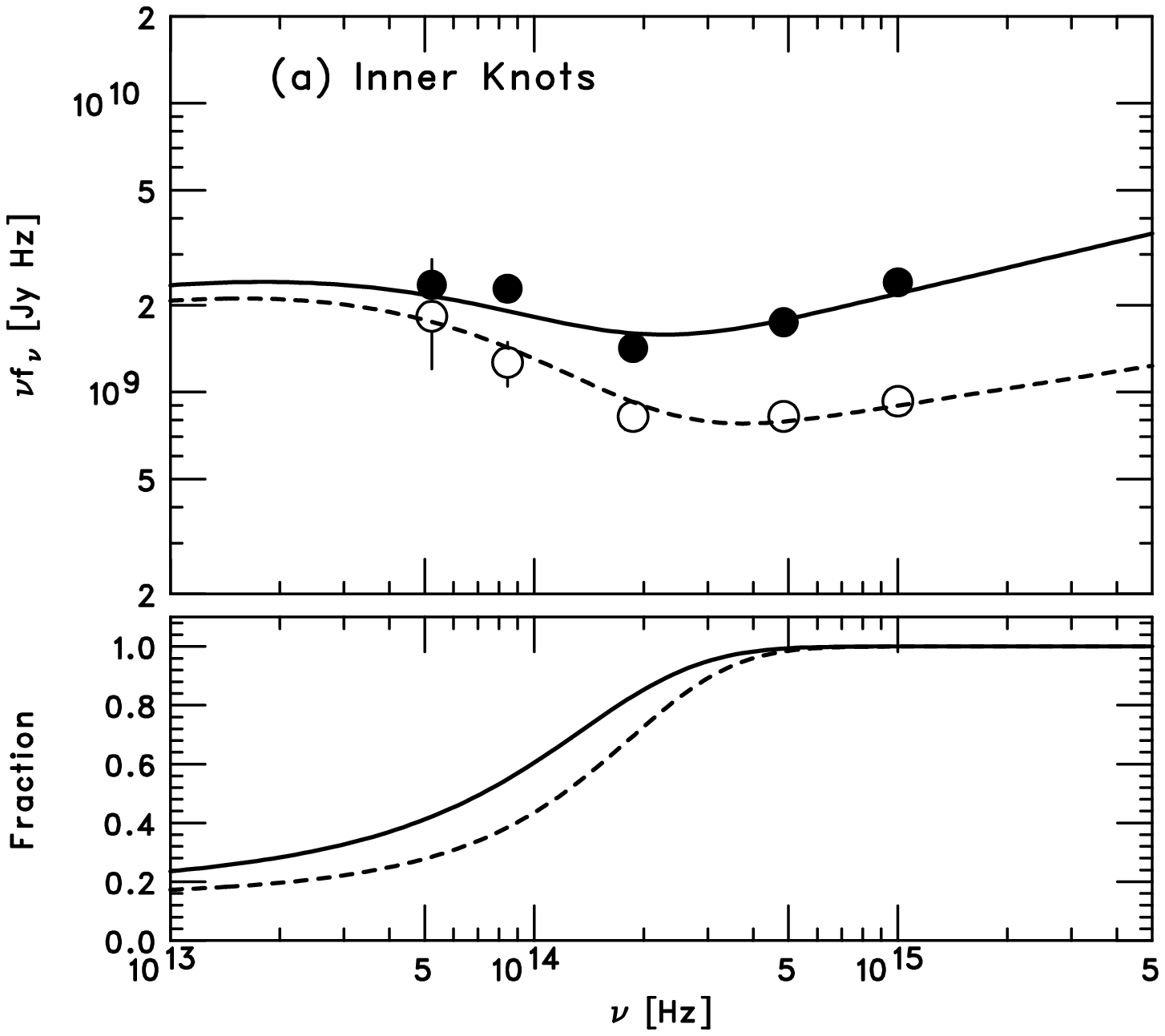}{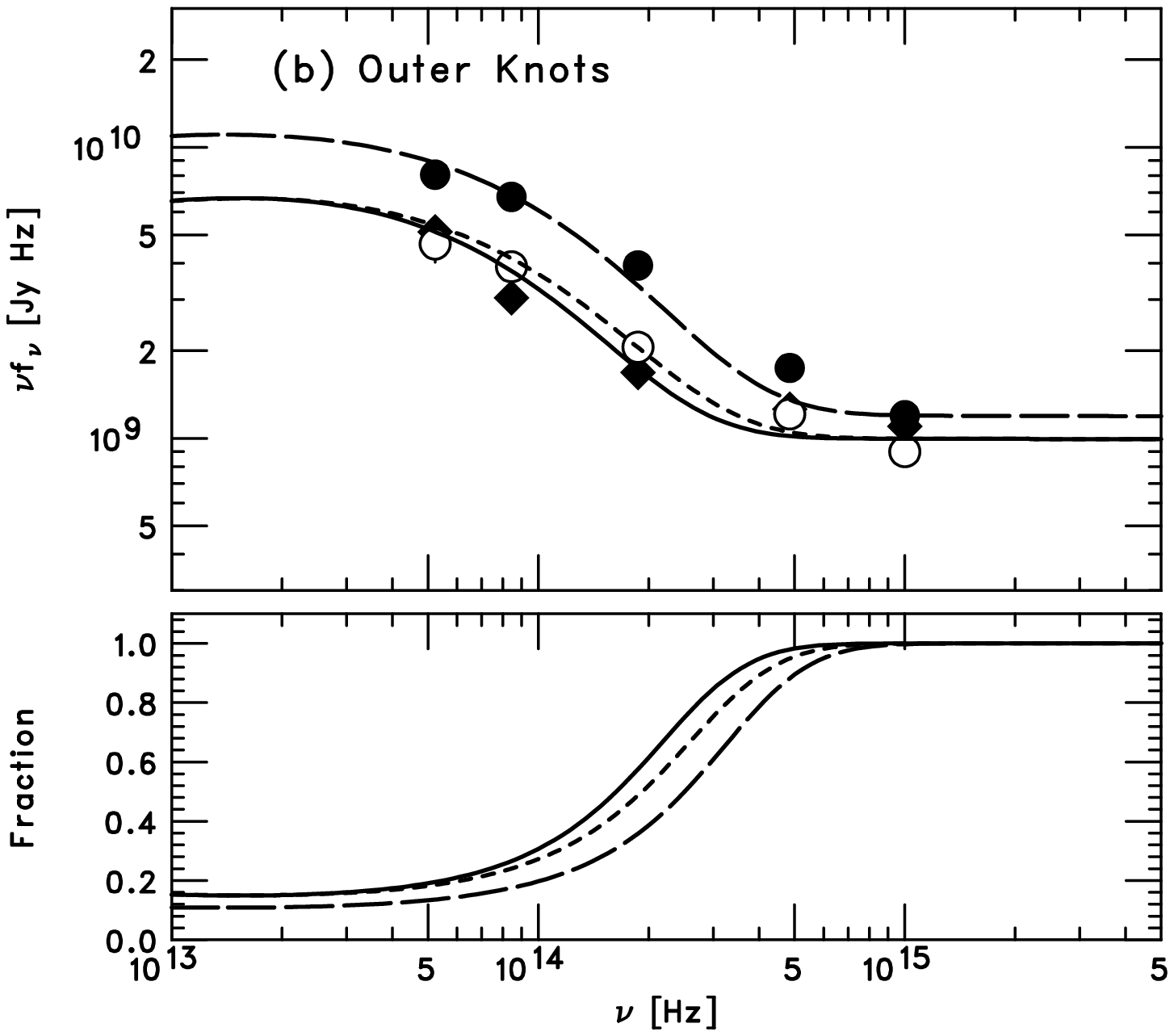}
\caption{({\it Top}) Close-up views (with same symbols) of the SEDs shown 
in Fig.~\ref{fig:SED} for the infrared-to-optical range.
({\it Bottom}) The contribution of the high-energy component 
relative to the total emission in terms of the two-component 
phenomenological model. 
\label{fig:IR-opt}}
\end{figure*}

In Fig.~\ref{fig:SED}, we show the SEDs  constructed for knots A, B1, C1, C2, and D1.
Since the interpretation of  multiple knots (B2/B3 and D2/H3) could be misleading, 
the ``summed" spectra for B2/B3 and D2/H3 have been omitted 
in  Fig.~\ref{fig:SED} and in the spectral modeling described below. 
The overall spectral appearance of the 
{\it inner knots} (A and B1) has strikingly different characteristics as compared to 
the {\it outer knots} (C1, C2, and D1). 
First, the {\it inner knots} radiate strongly in the X-ray (or even beyond)
with hard spectra 
while the {\it outer knots} radiate predominantly in the mid-infrared. 
Second, and most importantly, 
the SEDs of the {\it inner knots} have a concave shape from the mid-infrared to 
optical band,  indicating that 
two spectral components cross over at optical wavelengths.
Figure~\ref{fig:IR-opt} gives a close-up version of Fig.~\ref{fig:SED} 
highlighting the mid-infrared to near-ultraviolet part of the spectrum. 
We emphasize here that 
the IRAC fluxes  fill a  critical region  in the SEDs,  
making it possible to identify unambiguously the spectral structures present 
in the infrared to optical wavelengths.

In the SEDs of the {\it inner knots}, we  identify two spectral components, namely 
(1) the low-energy synchrotron spectrum extending from radio to infrared 
with a spectral cutoff at $\sim 5\times 10^{13}$ Hz, 
and (2) the high-energy component arising in the optical and smoothly 
connecting to the X-ray flux. 
Tentatively, we suggest that the 
high-energy component is  likely {\it of synchrotron origin}, because 
the optical polarization is consistent with the radio polarization 
in degree and orientation. 
In \S\S\ref{sec:syn2} and \ref{sec:IC} we discuss the origin of the high-energy spectral 
component and its implications to particle acceleration in the jet.
In explaining the broadband spectrum of knot A, 
a single power-law continuum all the way from radio to X-ray 
has been proposed in prior studies \citep{Roe00,Mar01}, but 
this picture does not correctly describe the present, more extensive data. 

In the {\it outer knots}, the SEDs also show the 
 two components but with different relative strengths. 
The low-energy part becomes more  dominant with increasing 
distance from the quasar core. 
At first glance the entire infrared-optical spectrum may be attributed to 
the falling part of synchrotron emission. 
However, careful inspection  shows a slight spectral-flattening 
in the  optical and near-ultraviolet, which suggests a contribution from 
the second component responsible for the X-ray emission, as 
already proposed by \citet{Jes02}. 
This idea is further strengthened  by the observation that 
the radio brightening in D1 (relative to C1/C2) 
accompanies infrared brightening 
while the near-ultraviolet and X-ray fluxes remain at similar levels
(see Fig.~\ref{fig:IR-opt}).

\begin{figure*}  
\plotone{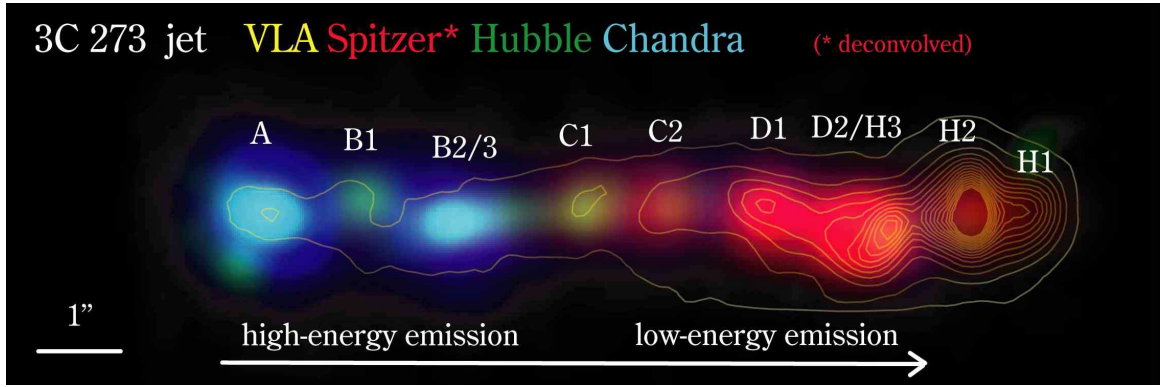}
\caption{\emph{Spitzer}-\emph{HST}-\emph{Chandra} composite image of
the jet of 3C~273. The colors are coded as follows: 
\emph{Spitzer} ``deconvolved" $3.6\ \mu\rm m$ (\emph{red}), 
\emph{HST} ``UV excess" $f_{0.3 \mu\rm m} - 0.05\times f_{1.6 \mu\rm m}$ (\emph{green}), 
\emph{Chandra} 0.4--6 keV (\emph{blue}). 
The \emph{VLA} radio (2 cm) contours are superposed on the image, 
with the strongest radio source H2 being truncated. 
There are two distinct types of radiating knots; the inner knots show 
hard optical spectra (\emph{green}) and strong X-rays (\emph{blue}) while 
the outer knots show bright infrared (\emph{red}). 
\label{fig:truecolor} }
\end{figure*}

To visualize  the spectral transition from the \emph{inner} 
to the \emph{outer} knots along the jet, 
we present a three-color image of the jet in Fig.~\ref{fig:truecolor} 
based on the data taken with 
\emph{Spitzer}, \emph{HST}, and \emph{Chandra}, 
with the \emph{VLA} contours from Fig.~\ref{fig:image} overlaid. 
The colors are coded as follows: 
\emph{Spitzer} $3.6\ \mu\rm m$ (\emph{red}), 
\emph{HST} $f_{0.3 \mu\rm m} - 0.05\times f_{1.6 \mu\rm m}$ (\emph{green}), 
\emph{Chandra} 0.4--6 keV (\emph{blue}). 
The \emph{Spitzer} image is a ``deconvolved"  IRAC image, 
which represents the results of our infrared photometry via 
PSF-fitting in \S\ref{sec:IRphoto}.
Specifically, the best-fit image obtained with two-dimensional fitting is 
artificially shrunk to match the effective resolution of the \emph{HST} image.
The \emph{HST} image is an ``UV excess" map 
[determined as $f_{0.3 \mu\rm m} - 0.05\times f_{1.6 \mu\rm m}$
based on the data in \citet{Jes05}]
representing 
the dominance of near-ultraviolet light over the near-infrared.
It is clear that the inner knots are luminous in both 
the UV excess and X-rays (the high-energy component), 
while the outer knots are bright in the mid-infrared (the low-energy component). 

\begin{table}[ht]
\begin{center}
\caption{Phenomenological Model Parameters of 3C~273 Jet Knots\label{tbl-2}}
\begin{tabular}{cccccccc}
\tableline\tableline
Parameter & A & B1& C1 & C2 & D1  & H2 & H1 \\
\tableline 
$\alpha_1$ & 0.76 & 0.73 & 0.63 &  0.68 & 0.75 & 0.9 & 0.9  \\
$\nu_c$ (Hz) & 
$5\times10^{13}$ & $5\times10^{13}$ & $4\times10^{13}$ & 
$5\times10^{13}$ & $6\times10^{13}$ &  
$5\times10^{13}$ &  $4\times10^{13}$ \\
$a$ & 0.8 & 0.8 & 0.8 &  0.8 & 0.8 & 0.8 & 0.8\\
$\alpha_2$ & 0.7 & 0.8 & 1.0 &  1.0 & 1.0 & \nodata & \nodata \\
\tableline
\end{tabular}
\tablecomments{ Equation (\ref{eq:fnu}), $f_{\nu} = 
\kappa_1 \nu^{-\alpha_1} \exp\left[ -\left( \nu/\nu_{c} \right)^{a}\right] + 
\kappa_2 \nu^{-\alpha_2}$, is adopted to characterize the SEDs.}
\end{center}
\end{table}

We describe the radio-to-X-ray spectra with the following double 
power-law models (one with an exponential cutoff): 
\begin{equation}
\label{eq:fnu}
f_{\nu} = f_{1\nu} + f_{2\nu} = 
\kappa_1 \nu^{-\alpha_1} \exp\left[ -\left( \nu/\nu_{c} \right)^{a}\right] + 
\kappa_2 \nu^{-\alpha_2}.
\end{equation}
The first term of the right-hand-side accounts for  the low-energy spectrum and 
the second pure power-law describes  the high-energy part.
(For the second component, $f_{2\nu} = \kappa_2 \nu^{-\alpha_2}$, 
we set an artificial low-energy turn-over below $10^{13}$ Hz.)
We introduced a parameter $a$ in the exponential part 
because particle acceleration models generally 
do not provide a robust prediction concerning the electron spectrum 
in the cutoff region. 
We adopt a common value of $a = 0.8$; the choice has little influence on other 
parameters. 
Table \ref{tbl-2} lists the spectral parameters that well 
describe the knot SEDs shown  in Fig.~\ref{fig:SED}.
Also, in Fig.~\ref{fig:IR-opt} we show the relative contribution of the 
second component, namely $f_{2\nu}/(f_{1\nu} + f_{2\nu})$, 
in the infrared to ultraviolet part. It should be noted that 
the second component is dominant ($>80\%$) 
already in the $R$ band flux and therefore the ``optical" jet traces 
the high-energy radiation. 
For knot A, the first component has a luminosity of 
$1.3\times 10^{43}\ \rm erg\ s^{-1}$ (in the frequency range 
$10^{9}\mbox{--}10^{14}$ Hz) without taking account of relativistic beaming. 
The minimum luminosity in the second component of knot A is 
$3.4\times 10^{43}\ \rm erg\ s^{-1}$ ($10^{14}\mbox{--}10^{18}$ Hz), 
and can be much higher if the spectrum extends well beyond the X-ray domain. 

The spectral index of the low-energy component is in the range of 
 $\alpha_1 = 0.63\mbox{--}0.76$,  as determined from the radio spectra. 
Remarkably, the cutoff frequency is quite similar among the knots, 
$\nu_c = (4\mbox{--}6)\times 10^{13}$ Hz, which differs 
from  previous modeling that deduced a progressive decrease of 
the cutoff frequencies \citep[see][and references therein]{Jes05}.
The difference in the deduced cutoff frequency results from 
our inclusion of the IRAC fluxes into the modeling. 
The second spectral index 
 for knots A and B1 is $\alpha_2=0.7\mbox{--}0.8$; 
for knots C1, C2 and D1,  the second index is steeper, $\alpha_2\simeq 1.0$. 
It should be emphasized that 
the second index of knot A agrees well with the local spectral index 
determined  by  X-rays alone, in other words  $\alpha_2 \simeq \alpha_{\rm x}$.
Interestingly, the steepness of the index seems to increase with 
decreasing X-ray intensity of the knots. 

There are some caveats for the interpretation of the SEDs that we constructed. 
As mentioned earlier, knot B1 in the optical shows a different morphology 
as compared to the radio one. Therefore, the first and second components 
should originate in different emission volumes.  
Also, the X-ray flux assigned for knot B1 could be contaminated by 
adjacent, bright knots. 

\subsection{Physical Parameters for the Low-Energy Component}
\label{sec:syn1}

The cutoff frequency of the low-energy synchrotron emission 
that we determined in the previous section 
represents the maximum energy of the associated electron population.
According to standard synchrotron theory \citep[e.g.,][]{GS65}, 
the maximum energy can be estimated as
\begin{equation}
\label{eq:emax}
E_{\max} \simeq 0.28\ 
\left[ \frac{1+z}{\delta} \right]^{0.5} 
\left( \frac{B}{0.1\ \rm mG}\right)^{-0.5} 
\left( \frac{\nu_{\rm c}}{10^{14}\ \rm Hz} \right)^{0.5} 
\ \rm TeV, 
\end{equation}
where $\delta$ is the Doppler factor, $z=0.158$ is the redshift, and 
$B$ is the comoving magnetic field strength.
The Doppler factor is defined as $\delta \equiv [\Gamma(1-\beta\cos\theta)]^{-1}$ 
with $\beta c$ the velocity of the jet, 
 $\Gamma = (1-\beta^2)^{-1/2}$ the bulk Lorentz factor of the jet, and 
$\theta$ the observing angle with respect to the jet direction. 
It is difficult to determine the Doppler factor, and therefore to what extent 
the observed jet radiation is enhanced by  Doppler beaming.
Furthermore, the jet may be decelerating along the observed jet portion 
resulting in a declining Doppler factor \citep{GK04,Tav06}.
A likely range is $1 < \delta < 10$ (see however \S\ref{sec:IC}).
If equipartition between the radiating electrons and 
magnetic fields is assumed, without taking into account any relativistic motion
(i.e., $\delta =1$), 
one can obtain an estimate of $B=0.1\mbox{--}0.2\ \rm mG$ \citep{Jes05}.
Therefore, given the cutoff frequencies of 
$\nu_c \simeq 5\times 10^{13}$ Hz, 
the maximum energy can be deduced as 
$E_{\rm max} \sim 0.2\ (B_{-4}\delta)^{-0.5}\ \rm  TeV$, 
where $B_{-4} = B/(0.1\ \rm mG)$.
According to the prescription given in \citet{HK02}, 
 the equipartition condition yields $B \propto \delta^{-1}$
 \citep[see,][for slightly different scaling]{Sta03}. 
 Then, an estimate of $E_{\rm max}$ is rather robust being nearly 
 independent of both $\delta$ and $B$. 

Synchrotron cooling in the source should affect the spectral shape provided that 
the cooling time is shorter than the formation timescale of the radiating knots. 
The synchrotron cooling time (in the jet comoving frame) for electrons 
with energy $E= \gamma m_e c^2$ 
emitting on average at a frequency $\nu$ (observed)  can be expressed as
\begin{eqnarray}
t_{\rm syn}
& = &  \frac{\gamma m_e c^2}{(4/3)c\sigma_{\rm T} U_B \gamma^2} , \\
& \simeq & 2400\, 
\left[ \frac{\delta}{1+z} \right]^{0.5} 
\left( \frac{B}{0.1\ \rm mG}\right)^{-1.5} 
\left( \frac{\nu}{10^{14}\ \rm Hz} \right)^{-0.5} 
 \rm yr, 
\end{eqnarray}
where $\sigma_{\rm T}$ is the Thomson cross section, and 
$U_B = B^2/(8\pi)$ is the energy density of the magnetic field.
Equation~(\ref{eq:emax}) was used in relating $E=\gamma m_e c^2$ with $\nu$, 
by replacing $E_{\max}$ with $E$ and $\nu_{\rm c}$ 
with $(\nu/0.29)$.
The source age, though quite uncertain, would be  $t_0 \sim 10^{5}\mbox{--}
10^{7}\ \rm yr$. 
Therefore, the observed synchrotron spectra from the radio to optical frequencies 
are likely to be formed in the 
  ``cooling regime", namely $t_{\rm syn} < t_0$,  which means that 
a continuous injection of electrons with an acceleration spectrum 
of $Q(E)\propto E^{-s_1}$ (in the power-law part) over the timescale $t_0$ 
gives rise to 
the energy distribution of synchrotron-emitting electrons 
$N(E)\propto t_{\rm syn} Q(E) \propto E^{-(s_1+1)}$.  
In this case,  according to the relation $s_1 = 2\alpha_1$, 
the observed index of $\alpha_1\simeq 0.7$ is translated 
into the electron acceleration index of $s_1\simeq 1.4$, which is coincident with 
the hardest possible index  ($s=1.5$) in the nonlinear 
shock acceleration theory \citep{MD01}. 
We implicitly assumed particle acceleration taking place in each knot \citep{Jes01},
where relativistic electrons are accumulated without escaping. 
A similar treatment was applied to the hotspot of 3C~273 
by \citet{MH86} to incorporate a shock-acceleration theory into the emerging 
radiation spectrum. 
There may be occasions for which the above argument is not applicable; 
for example, 
in the case of significant beaming, the synchrotron cooling time would be 
much longer ($t_{\rm syn} \propto \delta^{2}$ for the equipartition field) 
and the realization of the cooling regime would not be justified 
for radio-emitting electrons.

\subsection{Synchrotron Interpretation of the High-Energy Component}
\label{sec:syn2}

For the inner knots (A and B) we have found that the optical and X-ray fluxes 
can be modeled as a common power-law spectrum of spectral 
index $\alpha_2\simeq 0.7\mbox{--}0.8$. Also, the optical to X-ray 
slope agrees with the X-ray spectral index, especially for knot A, 
$\alpha_2 \simeq \alpha_{\rm x}$. 
One implication of this finding is that the optical and X-ray 
fluxes must be explained by the same emission mechanism. 
In \S\ref{sec:3c273}  we mentioned that 
both the radio and optical jet emission are  linearly polarized to 
 a similar degree of $\sim 15\%$ \citep{RM91,RCM96}, 
which has been taken as evidence for a synchrotron origin for the optical 
radiation in the inner knots. 
Now this in turn would imply a synchrotron origin for the X-ray radiation as well.
Therefore we first discuss the optical-X-ray component in terms of 
synchrotron radiation produced by high-energy electrons. 

According to the prescription in \S\ref{sec:syn1},  
the power-law slope of the energy distribution of the \emph{accelerated} 
electrons is $s_2 = 2\alpha_2$ in the synchrotron-cooling regime. 
Then the spectral index of the high-energy component, 
$\alpha_2\simeq 0.7\mbox{--}0.8$ for the inner knots, 
yields  $s_2\simeq 1.4\mbox{--}1.6$. 
Interestingly,  the index of acceleration is almost same between 
the low- and high-energy synchrotron components: 
$s_2 \sim s_1$ (because of  $\alpha_2 \sim \alpha_1$).  
On the other hand, the two  indices differ in  the outer knots: 
$\alpha_2\simeq 1$, corresponds to $s_2\simeq 2$.

The apparent difference in the cutoff frequency between the low- and 
high-energy components likely reflects a difference in 
the maximum energy of electron acceleration.  
The cutoff frequency associated with the high-energy component is not 
constrained by the current data, but must be greater than  $10^{18}$ Hz
($\sim 4$ keV). 
Consequently, from equation~(\ref{eq:emax}), the lower limit to the 
maximum electron energy is $E_{\rm max} \ga 30\ (B_{-4}\delta)^{-0.5}\ \rm  TeV$.

The presence of two different cutoff energies suggests 
two acceleration modes that can be characterized by 
different acceleration rates  in the knots. 
The acceleration timescale 
can be expressed as $t_{\rm acc} = \xi r_g/c$, 
with a gyroradius $r_g = E/(eB)$. 
The rate of acceleration is parametrized by 
the factor $\xi \geq 1$ which is, 
as frequently assumed, taken to be energy independent.
By equating $t_{\rm acc} = t_{\rm syn}$---namely, balancing the 
acceleration and synchrotron loss rates---one obtains the maximum 
attainable energy  limited by synchrotron losses:
\begin{equation}
E_{\max} \simeq 6\times 10^{15}\  \xi^{-0.5}  
\left( \frac{B}{0.1\ \rm mG}\right)^{-0.5} 
\ \rm eV. 
\end{equation}
Using equation~(\ref{eq:emax}) this 
can be translated into the corresponding cutoff frequency, 
\begin{equation}
\nu_{\rm c} \simeq 4.5 \times 10^{22} \ \xi^{-1} 
 \frac{\delta}{1+z} 
\ \rm Hz.
\end{equation}
Therefore, for the second synchrotron component to extend beyond $10^{18}$ Hz, 
the condition $\xi_2 < 4\times 10^4$ should be fulfilled. 
On the other hand, for the low-energy synchrotron component, 
 an extremely large value of $\xi_1 \sim 10^9$ can be inferred via a similar 
 argument. 
Interestingly, a study of multiwavelength spectra of small-scale blazar emission also 
gives  large $\xi$ \citep{IT96}.

In the theory of diffusive shock acceleration,  for relativistic shocks, 
the factor $\xi$ corresponds roughly to 
the ratio of the energy density of 
regular  magnetic fields to that of turbulent fields which resonantly scatter 
accelerating relativistic particles. 
Then a different value of $\xi$ means a different development of 
turbulent magnetic fields near the shock. 
The inferred large value of  $\xi_1$, following a discussion in \citet{Bie87}, 
 may be understood in terms of 
steep turbulence spectra, e.g., of Kolmogorov-type $I(k)\propto k^{-5/3}$
[where $I(k)$ is 
the magnetic energy density per wavenumber $k$ in the turbulent field], 
though it should be noted that $I(k)\propto k^{-1}$ was silently assumed above 
to obtain energy-independent $\xi$ for simplicity. 

There may be two distinct types of shocks characterized by 
different $\xi$ in the knots giving rise to their ``double-synchrotron" nature.
Alternatively, another acceleration mechanism 
may be operating, accounting  in a natural way for a different 
acceleration rate for the second component. 
For instance, turbulent acceleration may occur in the shear of 
jet boundary layers \citep{Ost00}, 
generating the second synchrotron component \citep{SO02}.
In highly turbulent shear layers, the acceleration rate can be estimated as 
$\xi  \sim  \beta_{\rm A}^{-2}$, where $\beta_{\rm A}$ denotes 
the Alfv\`en  velocity in units of $c$. Provided that the jet is composed of 
electron-proton plasma with the comoving number density $n$ \citep{SM00}, 
the Alfv\`en  velocity in the jet can be written as 
\begin{eqnarray}
\beta_{\rm A} &= & \frac{B}{\sqrt{4\pi nm_{\rm p}}} \\
& = & 7.3\times 10^{-2} 
\left( \frac{B}{0.1\ \rm mG}\right)
\left( \frac{n}{10^{-4}\ \rm cm^{-3}}\right)^{-0.5}  .
\end{eqnarray}
The condition $\xi_2 < 4\times 10^4$  needed to produce the 
X-rays can be fulfilled if $\beta_{\rm A} > 5\times 10^{-3}$, which seems 
reasonable for the typical jet parameters. 

An intriguing implication of interpreting the X-ray emission as being due to 
a synchrotron process would be the possibility that  protons in the jet 
 can be accelerated to very high energies.
In the case of protons, synchrotron losses do not limit the 
 acceleration process unless they  reach  ultra-high energies. 
Instead, a condition set by  $t_{\rm acc} = t_{\rm c}$,  where $t_{\rm c}$ is 
the  light-crossing timescale of the knot along the jet (typically $\sim 10$ kyr), 
would give a conservative estimate of the maximum attainable energy 
of protons.
Here we use a light-crossing time rather than the ``age" of the jet,  
probably $t_0 \sim 10^5\mbox{--}10^7\ \rm yr$, 
to conservatively account for an escape loss which takes at least longer than 
a light-crossing time.
This condition leads to 
\begin{eqnarray}
\label{eq:pmax}
E_{\rm p,max} &=& \xi^{-1}  ecB t_{\rm c} \\
&\simeq & 2.8\times 10^{20} \ \xi^{-1}  
\left( \frac{B}{0.1\ \rm mG}\right)
\left( \frac{t_{\rm c}}{10 \ \rm kyr}\right)
\ \rm eV, 
\end{eqnarray}
and consequently 
$E_{\rm p, max} > 0.7\times 10^{16}\ \rm eV$ for $\xi < 4\times 10^4$
and typical values of $B$ and $t_{\rm c}$. 
Remarkably, if $\xi \sim 10$ (close to the so-called Bohm diffusion
in the framework of diffusive shock acceleration), 
ultra-high-energy protons with energies 
$E_{\rm p, max} \sim  10^{19}\ \rm eV$ can be accelerated in the jet 
of 3C~273 \citep[see][for detailed theoretical arguments]{Aha02,HH04}. 
Even with a large value of $\xi$ ($\ga 100$), 
the shear acceleration 
would still be capable of generating $E_{\rm p, max} \sim  10^{19}\ \rm eV$, 
since the turbulent acceleration in the shear layer 
may be available for much longer timscales. 
The fraction of jet power that goes to high energy protons 
depends on the particle content of the jet and the details of the physics of 
particle acceleration, both of which are not well known.  

It is interesting to note that if protons are indeed accelerated in the jet 
to very high energies, $E_{\rm p, max} \ga  10^{18}\ \rm eV$, 
the second synchrotron component can be produced directly by \emph{protons}, 
i.e., proton synchrotron radiation \citep{Aha02}, rather than by electrons. 
For the same particle energy, the characteristic frequency of 
proton synchrotron radiation, $\nu_{\rm p,c}$, is less than 
that of electron synchrotron radiation by a factor of $(m_{\rm e}/m_{\rm p})^3$. 
Therefore, by using equation~(\ref{eq:emax}), the spectrum of proton synchrotron 
has a cutoff frequency of 
\begin{equation}
\nu_{\rm p,c} \simeq 2.1 \times 10^{18} \ 
\left[ \frac{\delta}{1+z} \right]
\left( \frac{B}{\rm mG}\right) 
\left( \frac{E_{\rm p, max}}{10^{18}\ \rm eV} \right)^2\ 
\rm Hz, 
\end{equation}
which is located above the X-ray domain.
Thus protons rather than a second population of electrons, may 
explain the  second synchrotron component. 
Unlike the electron-synchrotron case, 
the diffuse nature of the emission (at least in the optical) 
can be understood more easily  thanks to much longer cooling timescale for protons. 
A magnetic field strength of $\ga 1\ \rm mG$ is needed 
in order to keep   the power spent to accelerate protons at 
a reasonable level, and at the same time to render the acceleration 
timescale shorter and the diffusion timescale longer \citep{Aha02}.
By adopting $B=10\ \rm mG$, $\delta=1$, 
an escape loss timescale of 
$t_{\rm esc}= 1.4\times 10^7(E_{\rm p}/10^{14}\ \rm eV)^{-1/2}\ yr$, 
and a continuous proton injection over $t_0 = 3\times 10^7\ \rm  yr$, 
\citet{Aha02} obtained the acceleration power for protons 
of $L_{\rm p} \simeq 10^{46}\ \rm erg\ s^{-1}$ to explain 
the optical-X-ray spectrum of knot A in 3C~273 in terms of 
proton synchrotron radiation.

\subsection{Beamed IC Model for the High-Energy Component}
\label{sec:IC}

Previous explanations of the second, X-ray-dominated component of the SED 
for jet knots have emphasized 
the beamed IC scenario \citep{Tav00,CGC01}.
Specifically,  relativistically beamed IC emission in the CMB photon field 
has been invoked to model the X-ray emission in 3C~273 \citep{Sam01}, 
as well as in other quasar jets. 
This model requires a highly relativistic bulk velocity of the jet, 
with a Doppler factor 
$\delta \sim 10$ out to nearly megaparsec distances. 
A direct connection between the optical and X-ray fluxes 
in the inner knots A and B1 and also between the ultraviolet and X-ray fluxes 
in the outer knots (see Fig.~\ref{fig:SED})
 led us to consider that the same mechanism should 
be responsible for the optical and X-ray emissions, forming a distinct high-energy 
component. 
If one adopts the beamed IC model for the X-rays, 
the optical fluxes in the inner knots and the ultraviolet fluxes 
in the outer knots  should be explained also by the beamed IC 
radiation. This implies  that 
the energy distribution of electrons has to continue down to very 
low energies of $E_{\rm min} \sim m_e c^2$ without a cutoff or break.

In this scenario, one still should account for the optical polarization. 
In the case of  ``cold" electrons in the jet, 
Compton up-scattering off the CMB by an ultrarelativistic jet 
(so-called  ``bulk Comptonization") 
can yield in principle  a  high degree of polarization 
for the standard choice of $\theta \simeq \Gamma^{-1}$ \citep{BS87}. 
However, for relativistic electrons with $E \sim 2\, m_ec^2$ 
as appropriate for the optical IC emission in this jet, 
the degree of polarization should be significantly reduced \citep{Pou94}. 
The present data of the radio and optical 
polarization \citep{RCM96} may require a tuned set of physical conditions 
 to reconcile with the beamed IC scenario. 

In this respect, a possible caveat would be 
that in addition to the CMB, the synchrotron radiation 
of the jet itself might serve as seed photons for Compton up-scattering;  
if the highly relativistic jet accompanies slowly moving outer portions, 
the synchrotron radiation that comes from the slowing part can be 
relativistically enhanced in the fast moving jet, providing the effective 
photon fields for Comptonization \citep[][as discussed for small-scale jets]{GTC05}.
If the observed radio flux is largely contributed by a slow,  
non-relativistic layer, 
the energy density of the synchrotron 
photons (as measured in the observer's frame) could reach 
$\sim 10^{-12}\ \rm erg\ cm^{-3}$, being comparable to the CMB energy density.
This possibility introduces a complication to the polarization argument. 

Within the framework of the beamed IC radiation, the physical parameters 
that can reproduce the SEDs of the 3C~273 jet 
 have been obtained \citep[e.g.,][]{Sam01,HK02}.
Compared to a typical value of 
 $E_{\rm min} \sim 20\, m_e c^2$ in the previous work \citep[e.g.,][]{Sam01}, 
a smaller minimum energy of the electron distributions, say 
 $E_{\rm min} \sim m_e c^2$, 
is required by the present data. 
Note that this increases the estimate of the jet kinetic power by an order of 
magnitude if the jet power is dynamically dominated by cold protons. 
Let us model the SED of knot A with a synchrotron plus beamed IC model of 
\citet{Tav00} assuming $\delta = \Gamma$ 
and  a knot radius of 1 kpc (in the jet frame of reference). 
For a Doppler factor of 
$\delta = 30$ and magnetic field $B=5\ \mu$G as roughly corresponding 
to an equipartition condition, 
we obtain a jet kinetic power of 
$L_{\rm jet} \simeq 1\times 10^{48}\ \rm erg\ s^{-1}$, 
if the number of the observed electrons is balanced by the number of 
cold protons in the jet.
However, such a large Doppler factor of 
$\delta = 30$  seems problematic, 
since relatively large contributions from 
the un-beamed components are observed in the core emission,  
such as an ultraviolet bump 
\citep[see][and references therein]{Cou98}, and the 
accretion-disk emission comparable to the jet emission
in the X-ray regime \citep{GP04}.  
Instead, if we adopt a smaller  Doppler factor of $\delta =10$ 
(together with $E_{\rm min} = 3\, m_e c^2$ to account for the 
optical emission),  
the energy density of electrons exceeds that of magnetic fields 
 more than three orders of magnitude, and the jet kinetic power reaches an 
uncomfortable level of $L_{\rm jet} \simeq 1\times 10^{49}\ \rm erg\ s^{-1}$.
 To summarize, the beamed IC model, in its simple form, 
does not easily find a satisfactory set of parameters, 
particularly for the case of knot A.

Finally, the upper limit to the infrared fluxes from the quiet portion of the jet 
upstream of knot A is important to constrain the structure of the 
jet in the context of the beamed IC model \citep{Geo05,Uch05}. 
Unfortunately, we need additional IRAC observations to 
derive a reliable upper limit there because of the large systematic 
uncertainty in PSF subtraction in the region close to the quasar core.

\subsection{The Head Region}

\begin{figure}
\epsscale{1.0}
\plotone{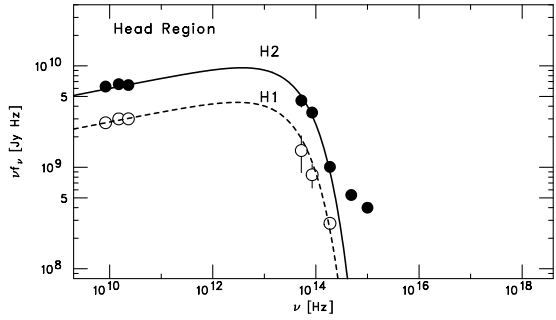}
\caption{Broadband SEDs of the head region of the 3C~273 jet:
H2 (filled circles) and H1 (open circles). 
The photometric data are given in Table \ref{tbl-1}. 
The curves represent the single-component model  (see the text).  \label{fig:SED2}}
\end{figure}

In Fig.~\ref{fig:SED2}, we show the SEDs obtained for H1 and H2 in 
the head region.  Note that no significant X-ray emission was found in this 
region. 
We have modeled the SEDs  using a power law with an exponential-type cutoff, 
namely by equation~(\ref{eq:fnu}) with $\kappa_2=0$, as drawn in Fig.~\ref{fig:SED2}.
The spectral parameters used for the model are indicated 
in Table \ref{tbl-2}.  Interestingly, the spectrum of H2 shows a notable flattening 
at the optical and ultraviolet wavelengths, as first reported by \citet{Jes05}. 

The radio spectra are steeper than those of the upstream knots, with 
$\alpha \simeq 0.9$. 
It is likely that particle acceleration acting in the head region 
would be terminal shocks that 
have different properties as compared to  upstream shocks.  
As argued by \citet{MH86}, 
if the effect of synchrotron cooling is taken into account, the index seems 
consistent with an acceleration spectrum of $Q\propto E^{-2}$, which is 
 the standard spectrum of diffusive shock acceleration.
Within the picture proposed by \citet{MH86}, however, the excess fluxes 
 at optical and ultraviolet wavelengths seen in the spectrum of H2 cannot 
 be easily understood. These excesses may represent a second synchrotron 
 component just as in the case of the upstream knots.

\section{Conclusions} 
\label{sec:conclusions}

We have presented results from our \emph{Spitzer} IRAC 
observation of the 3C~273 jet at wavelengths 3.6 and $5.8\ \mu\rm m$, 
combined with the photometry with the 
\emph{VLA} radio,  \emph{HST} optical, and \emph{Chandra} X-ray data. 
Our multiwavelength analysis led us to conclude that 
the flat optical emission in the X-ray-dominated knots 
originates in the high-energy power-law component, which also accounts for the X-ray 
emission. 
The agreement between the optical-X-ray slope ($\alpha_2$) and 
the X-ray spectral index ($\alpha_{\rm x}$) throughout the jet supports this picture. 
On the other hand, the radio to infrared spectra can be 
expressed by a power law with an exponential cutoff at $\sim 5\times 10^{13}$ Hz. 
The two distinct radiation components, 
namely low-energy (radio-infrared) and high-energy (optical-X-ray) emission, 
 have similar power of $\sim 10^{44}\ \rm erg\ s^{-1}$ for the entire  jet volume 
(without a beaming correction); the power in the second component becomes
noticeably higher  if its peak position is located far beyond the X-ray domain. 
The relative importance of the two components changes along the 
jet (see Fig.~\ref{fig:truecolor}). 
In the inner, X-ray-dominated knots, the high-energy component
overwhelms the low-energy one. 

The second component can be attributed to either synchrotron radiation by 
a second population of high-energy electrons (or protons), or the beamed IC emission 
by the radio-emitting electrons. 
In the first case, the double  synchrotron nature  may arise from the presence of 
distinctively different acceleration processes (e.g., shock and 
turbulent acceleration).  
In the context of the origin of extragalactic cosmic-rays, it is interesting 
that a faster acceleration mechanism producing the second 
component is capable of  accelerating cosmic-ray protons to energies 
$10^{16}\ \rm eV \la E_{\rm p, max} \la  10^{19}\ \rm eV$.
On the Comptonization interpretation of the high-energy emission,  
we argued that the present polarization data seems problematic. 
Future sensitive measurements of polarization are quite 
important to draw firmer conclusions about the origin of the high-energy emission.

The identification of the synchrotron component peaking at 
$\sim 5\times 10^{13}\ \rm Hz$ in the X-ray-dominated knots
of the 3C~273 jets calls for a follow-up deep observation with the \emph{Spitzer} IRAC with better configurations.   
Our \emph{Spitzer} cycle-3 proposal (PI: Uchiyama) 
requesting follow-up deep observations of 3C~273, as well as 
the observations of 10 other jets, has recently been approved. 
Ultimately, 
the \emph{James Webb Space Telescope}, 
a future large infrared-optimized space  telescope, 
will have an excellent capability to explore the jet emission in the IRAC band 
with much improved resolution and sensitivity.  

An interesting implication of this work is that a number of optical jets 
in radio-loud quasars discovered so far may be directly related 
to the X-ray emission rather than to the radio emission, as  in the 3C~273 jet. 
For example, in \citet{Uch05}, we have associated the optical flux 
of the jet knot in the quasar PKS 0637$-$752 with 
the highest energy part of the synchrotron component. 
However, the optical flux could be attributed equally to the 
low energy part of the second component, given the result presented here. 
This issue should be addressed in future analyses 
based on multiwavelength observations of \emph{Chandra}-detected quasar jets. 
[In this context, we note that our discussions 
are not concerned with 
 low-luminosity FR I jets such as M~87 \citep[e.g.,][]{PW05}, in which 
the X-ray emission appears to lie on the extrapolation of the emission at lower frequencies, and is widely believed to be of synchrotron origin.]
It is clear from our results that the measurements over as wide 
a wavelength coverage as possible in the infrared-optical domain 
are of great importance 
in studying the particle acceleration  mechanisms 
in the large-scale jets of powerful quasars.
Unfortunately, quasar jet emission in the optical region of the spectrum 
is very faint, and only a few optical slopes for the powerful quasars 
have been measured \citep{Rid97,Che02}. 
Further \emph{Spitzer} observations are therefore particularly helpful 
even though the spatial resolution is not optimal for the study of jet emissions. 
Also, a careful measurement of X-ray spectral index and its comparison 
with the optical fluxes is valuable to 
progress toward the understanding of the dominant 
X-ray emission mechanism operating in the quasar jets.


\acknowledgments

We wish to thank Felix Aharonian for invaluable discussion.
We also thank the anonymous referee for constructive comments 
and suggestions.
This work was supported in part by NASA grant NAG5-12873.
This work is based on observations made with the Spitzer Space Telescope,
which is operated by the Jet Propulsion Laboratory, California Institute of 
Technology under NASA contract 1407. Support for this work was provided 
by NASA through Contract Number RSA 1265389 issued by JPL/Caltech.
The National Radio Astronomy Observatory is operated by Associated
Universities, Inc.\ under a cooperative agreement with the National Science
Foundation.

\end{document}